\documentclass[aps,prl,reprint,groupedaddress]{revtex4-2}

\usepackage{graphicx}
\usepackage{amsmath}

\begin{document}


\title{Recovery of tunable bound states in the continuum}




\author{Hai~Huang}
\author{Huiming~Zhang}
\author{Biye~Xie}
\author{Wengang~Bi}
\email[]{waynebi@cuhk.edu.cn}
\author{Daquan~Zhang}
\email[]{zhangdaquan@cuhk.edu.cn}
\affiliation{%
	School of Science and Engineering, The Chinese University of Hong Kong (Shenzhen), Longgang, Shenzhen, Guangdong, 518172, P.R. China.
}%

\date{\today}

\begin{abstract}
	Tunable bound states in the continuum (BICs) in photonic crystal slabs are highly sensitive to substrate-induced mirror-symmetry breaking and typically degrade into finite-$Q$ quasi-BICs in realistic integrated platforms. Here we show that such degradation can be deterministically reversed. Using temporal coupled-mode theory and full-wave simulations, we demonstrate that the radiation channel opened by the substrate can be exactly cancelled by introducing a second, independent odd-parity perturbation inside the slab. This dual-asymmetry strategy restores the singularity of the radiation matrix and thereby recovers a tunable BIC in a substrate-supported photonic crystal slab. The recovered state regains both the polarization vortex and the characteristic $Q\propto \Delta k^{-2}$ scaling. The recovery points further follow a linear relation in the two-asymmetry parameter space, revealing a simple mode-dependent compensation law. The same mechanism also restores merging-BIC configurations, showing that it applies not only to isolated tunable BICs but also to higher-order topological resonance states built from them. Our results establish a practical route for preserving tunable topological resonances in substrate-supported nanophotonic systems.
\end{abstract}


\maketitle



Bound states in the continuum (BICs) provide a powerful route to confine waves without radiative loss even though the corresponding eigenstates lie inside the spectrum of outgoing channels~\cite{friedrich_Interfering_1985,hsu_Bound_2016}. In photonic crystal slabs, they have evolved from a theoretical concept into a practical design principle for realizing extremely high-$Q$ resonances in open nanophotonic systems~\cite{lee_Observation_2012,hsu_Observation_2013,yang_Analytical_2014,suh_Temporal_2004}. Optical BICs may arise from symmetry incompatibility or from destructive interference between leakage pathways, and these perspectives together provide a useful framework for understanding guided resonances in periodic slabs~\cite{suh_Temporal_2004,yang_Analytical_2014,gao2016formation,blanchard2016fano}. Beyond isolated nonradiative states, BICs in photonic crystal slabs are now understood to possess a momentum-space topological nature. They are associated with polarization singularities in the far field and carry quantized topological charges that govern the generation, evolution, and annihilation of high-$Q$ resonances~\cite{zhen_Topological_2014,bulgakov2017bound,doeleman2018experimental,hu_Global_2022}.

Among the different classes of optical BICs, accidental, i.e.\ tunable, BICs are of particular interest because they occur away from the $\Gamma$ point and can be moved, created, or merged through continuous variation of structural parameters~\cite{yang_Analytical_2014,ni2016tunable,hu_Global_2022}. This tunability makes them attractive for shaping the momentum-space distribution of radiative loss and for constructing more complex topological resonance configurations, including merging-BIC states with unusually broad high-$Q$ regions~\cite{kang2021merging,jin_Topologically_2019}. However, the same interference mechanism that gives tunable BICs their flexibility also makes them fragile. Even modest structural or environmental asymmetry can break the radiation cancellation condition and convert an ideal BIC into a finite-$Q$ quasi-BIC~\cite{ni2016tunable,ovcharenko_Bound_2020}.

This fragility is particularly limiting in integrated nanophotonics. In most experimentally relevant slab geometries, the photonic crystal must be supported by a substrate, which breaks the up-down mirror symmetry and opens an additional radiation channel with odd-parity~\cite{ovcharenko_Bound_2020,cerjan2019bound}. As a result, a tunable BIC generally collapses into a leaky resonance, its polarization singularity is modified, and the ideal divergence of the $Q$ factor is lost. Previous studies have broadened the design space of photonic BICs through anisotropy engineering, environmental design, double-layer structures, multipolar analysis, and topological merging~\cite{gomis2017anisotropy,li2016doublelayer,cerjan2019bound,sadrieva2019multipolar,kang2021merging}. These efforts have greatly advanced the understanding of how BICs emerge and evolve in open systems, and have enabled applications ranging from BIC lasers to strongly enhanced nonlinear optical processes~\cite{kodigala2017lasing,carletti2018giant,koshelev2020subwavelength,hwang2022nanophotonic,kang2023applications}. Several recent works have proposed strategies to recover BICs in substrate-supported photonic crystal slabs, including using tapered holes and depositing a thin film on the slab surface~\cite{xu_Tapered_2024,bai_Recovery_2025}. However, a deterministic strategy for recovering a substrate-degraded tunable BIC in an otherwise conventional supported photonic crystal slab has not yet been established.

Here we show that the degradation of substrate-supported tunable BICs can be reversed in a controlled and predictive manner. Starting from temporal coupled-mode theory, we show that the radiation channel opened by substrate-induced out-of-plane asymmetry can be cancelled by a second, independent odd-parity perturbation inside the slab. This compensation restores the singularity of the radiation matrix and thereby recovers an exact tunable BIC in a substrate-supported photonic crystal slab. Our formulation identifies the common odd-parity origin of the compensation mechanism and predicts a mode-dependent linear recovery condition in the two-asymmetry parameter space. Full-wave simulations of a silicon photonic crystal slab confirm that the recovered state regains both the polarization vortex and the characteristic $Q\propto \Delta k^{-2}$ scaling. We further show that the same dual-asymmetry compensation principle also applies to merging-BIC configurations, indicating that the mechanism is not limited to isolated tunable BICs.


\begin{figure*}
	\centering
	\includegraphics[width=1\textwidth]{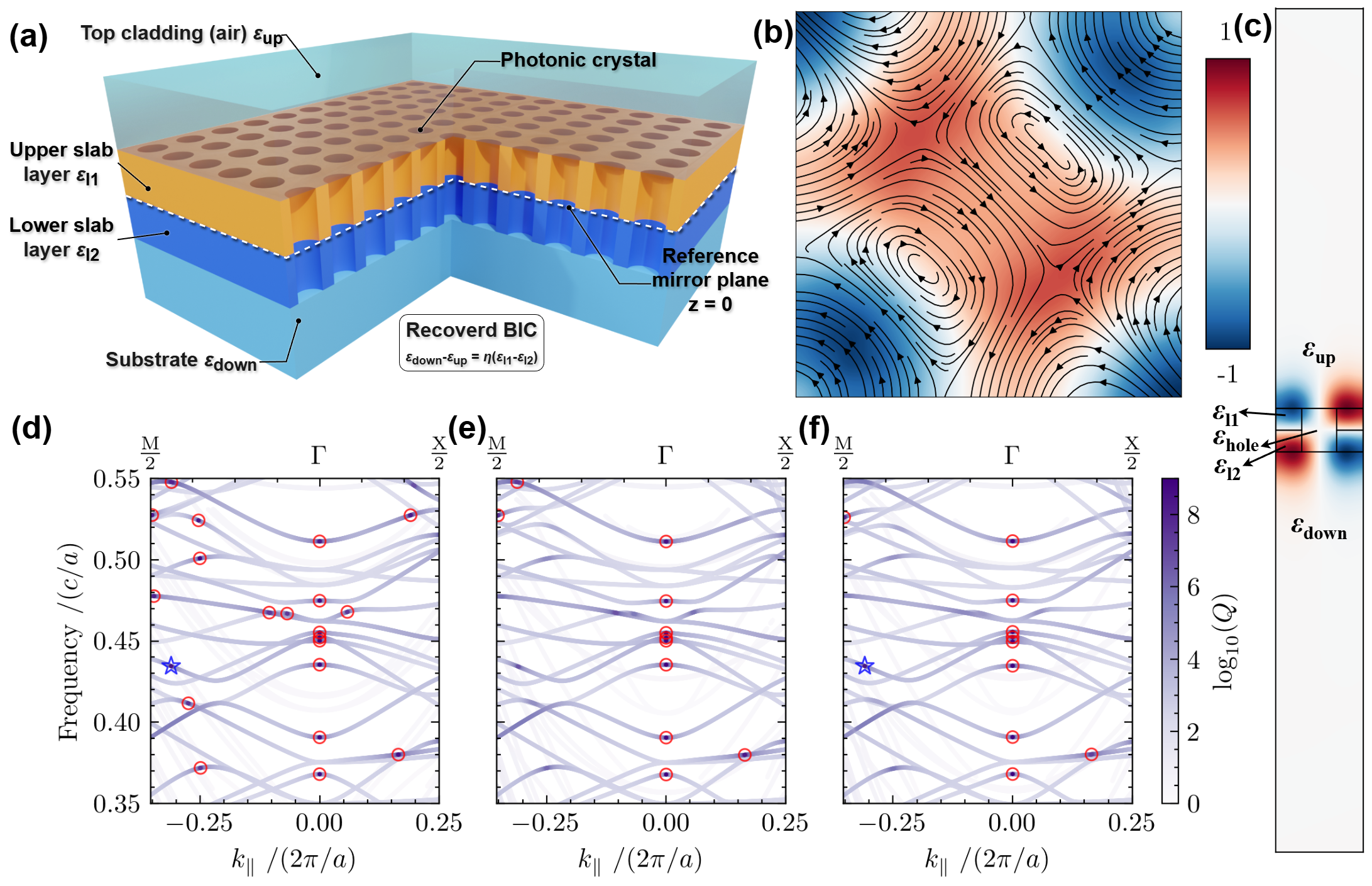} 

	\caption{Mechanism of BIC degradation and recovery.
		(a) Illustration of the method for recovering tunable BICs in substrate-supported photonic crystals.
		(b) and (c) Electromagnetic field distribution of the tracked BIC mode in the (b) $x$-$y$ and (c) $y$-$z$ plane, respectively. The color represents the mode profile of $E_z$ for (b) and $E_x$ for (c). The streamlines in (b) represent the electric field vector.
		(d)--(f) Band structures of the photonic crystal slab for different perturbation parameters:
		(d) $(\Delta \varepsilon_1,\Delta \varepsilon_2)=(0,0)$,
		(e) $(0.5368,0)$, and
		(f) $(0.5368,0.7999)$.
		The color of the band structure represents the $\log_{10}(Q)$ factor of the modes. The tracked BIC mode is marked by blue star. Other BIC or quasi-BIC modes with $Q>10^7$ are marked by red circles.
	}
	\label{fig:1-band} 
\end{figure*}

We begin from a two-mode description of Friedrich-Wintgen BIC\cite{friedrich_Interfering_1985}. In temporal coupled-mode theory, the system can be described by a non-Hermitian Hamiltonian
\begin{equation}
	H = \mathbf{\Omega}-i\mathbf{\Gamma}
	=\begin{pmatrix}
		\omega_1 & \kappa   \\
		\kappa   & \omega_2
	\end{pmatrix}
	-
	i\begin{pmatrix}
		\gamma_1   & \gamma_0 \\
		\gamma_0^* & \gamma_2
	\end{pmatrix},
\end{equation}
where $\omega_i$ is the resonant frequency, $\kappa$ is the near-field coupling coefficient, $\gamma_i$ is the decay rate, and $\gamma_0$ is the far-field coupling coefficient between the two modes.
For a BIC to exist, the eigenstate $|\psi_{\mathrm{BIC}}\rangle$ must possess a purely real eigenvalue, which inherently demands that the expectation value of the dissipation matrix vanishes: $\langle \psi_{\mathrm{BIC}} | \mathbf{\Gamma} | \psi_{\mathrm{BIC}}\rangle = 0$. Because $\mathbf{\Gamma}$ is a positive semi-definite matrix, this equates to the singularity condition $\det(\mathbf{\Gamma}) = 0$.
We use the radiation coupling matrix $\mathbf{D}$ to represent the coupling between the modes and the radiation channels, which is
\begin{equation}
	\mathbf{D} =
	\begin{pmatrix}
		d_{1}^+ & d_{2}^+ \\
		d_{1}^- & d_{2}^- \\
	\end{pmatrix},
\end{equation}
where $d_{i}^+$ and $d_{i}^-$ are the coupling coefficients of the $i$-th mode to the radiation channels in the $+z$ and $-z$ directions, respectively. $2\mathbf{\Gamma}=\mathbf{D}^\dagger\mathbf{D}$ can be derived from energy conservation, which means $\mathbf{\Gamma}$ and $\mathbf{D}$ share the same singularity condition~\cite{suh_Temporal_2004}.
Thus the necessary condition for a BIC reduces directly to the singularity of the radiation matrix:
\begin{equation}
	\det(\mathbf{D}) = d_1^+ d_2^- - d_1^- d_2^+ = 0.\label{eq:det-d}
\end{equation}
In systems with out-of-plane $\sigma_z$ symmetry, this condition is intrinsically satisfied.

Tunable BICs typically emerge from the coupling of guided resonances and Fabry-Perot modes that share the same $\sigma_z$ parity ($p_0 = \pm 1$)~\cite{hu_Global_2022}. However, this delicate radiation balance is highly sensitive to asymmetric perturbations. A prototypical scenario is the introduction of a substrate. For mathematical simplicity, we redefine the background $\varepsilon_0(\mathbf{r})$ to preserve strict $\sigma_z$ symmetry, treating the asymmetric claddings entirely as an odd-parity perturbation $\hat{V}_a(z)$ with intensity $\lambda_a$ and parity $p_a = -1$.

According to first-order perturbation theory, the radiation amplitudes evolve as $d_i^\pm = d_i^{\pm(0)} + c_{ia}^\pm \lambda_a$, where $c_{ia}^\pm$ are proportionality constants determined by the mode profiles. Spatial symmetry dictates that the unperturbed amplitudes satisfy $d_i^{-(0)} = p_0 d_i^{+(0)}$, while the odd-parity perturbation enforces $c_{ia}^- = p_a p_0 c_{ia}^+ = -p_0 c_{ia}^+$. The symmetry relations for the unperturbed and first-order radiation amplitudes are derived rigorously in the Supplemental Material~\cite{SI}, Sec.~S2.2. Substituting these symmetry constraints into the determinant yields:
\begin{equation}
	\det(\mathbf{D}) = 2p_0 \left( d_2^{+(0)}c_{1a}^+ - d_1^{+(0)}c_{2a}^+ \right) \lambda_a.\label{eq:det-d-a}
\end{equation}
Because all other terms exactly vanish within the first-order perturbation framework, $\det(\mathbf{D})$ exhibits a strict linear dependence on $\lambda_a$. Consequently, the introduction of the substrate immediately breaks the singularity condition, inevitably degrading the BIC into a quasi-BIC with a finite $Q$ factor.

It naturally follows that the singularity of the radiation matrix can be recovered by compensating the system with an additional linear perturbation term. Accordingly, we introduce a secondary, independent odd-parity perturbation $\hat{V}_b(z)$ with intensity $\lambda_b$ and radiation corrections $c_{ib}^\pm$. Under the influence of both perturbations, the first-order radiation amplitudes evolve as $d_i^\pm = d_i^{\pm(0)} + c_{ia}^\pm \lambda_a + c_{ib}^\pm \lambda_b$. Applying the spatial parity constraints to this dual-perturbation framework, the determinant expands to:
\begin{equation}
	\begin{aligned}
		\det(\mathbf{D}) & =  2p_0 \left( d_2^{+(0)}c_{1a}^+ - d_1^{+(0)}c_{2a}^+ \right)\lambda_a   \\
		                 & \; + 2p_0 \left( d_2^{+(0)}c_{1b}^+ - d_1^{+(0)}c_{2b}^+ \right)\lambda_b
	\end{aligned}\label{eq:det-d-ab}
\end{equation}
Consequently, tuning the secondary parameter $\lambda_b$ allows us to explicitly enforce $\det(\mathbf{D}) = 0$. This establishes a deterministic, linear correlation between $\lambda_a$ and $\lambda_b$ that cancels the substrate-induced radiation loss, ensuring the exact topological recovery of the tunable BIC.


To verify this mechanism, we consider a square-lattice photonic crystal with lattice constant $a$, using Si ($n=3.77$)~\cite{karaman_Decoupling_2025} as the photonic crystal slab, air ($n=\sqrt{\varepsilon_{\mathrm{up}}}=1$) as the top cladding and silica ($n=\sqrt{\varepsilon_{\mathrm{down}}}=1.44$)~\cite{malitson_Interspecimen_1965} as the bottom cladding. To simplify the discussion, frequencies are normalized by $c/a$, where $c$ is the speed of light in vacuum, and wave vector by $2\pi/a$. We set the radius of the air hole to be $0.20a$ and the thickness of the slab to be $0.50a$.

First, we consider the ideal case which has a $\sigma_z$ symmetric cladding with permittivity
\begin{equation}
	\varepsilon_{\sigma_z}=\frac{1}{2}(\varepsilon_{\mathrm{up}}+\varepsilon_{\mathrm{down}}).
\end{equation}
Because of the symmetries $C_{2}^{z}T$ and $\sigma_{z}$, stable BICs can exist in momentum space~\cite{zhen_Topological_2014}. We calculate the band structure of the photonic crystal slab using finite element method (FEM) and mark the possible BICs ($Q>10^7$) in Fig.~\ref{fig:1-band}(d)-(f). Here we choose $(k_{x},k_{y})=(-0.219,-0.219)$ as the tracked BIC mode (blue star in Fig.~\ref{fig:1-band}(d)), which has a normalized frequency of $0.435$. This mode lies away from the $\Gamma$ point and therefore is not symmetry-protected. The mode profiles of the tracked BIC are shown in Fig.~\ref{fig:1-band}(b) and (c) for $z=0$ and $x=0$ planes, respectively.

\begin{figure}
	\centering
	\includegraphics[width=0.48\textwidth]{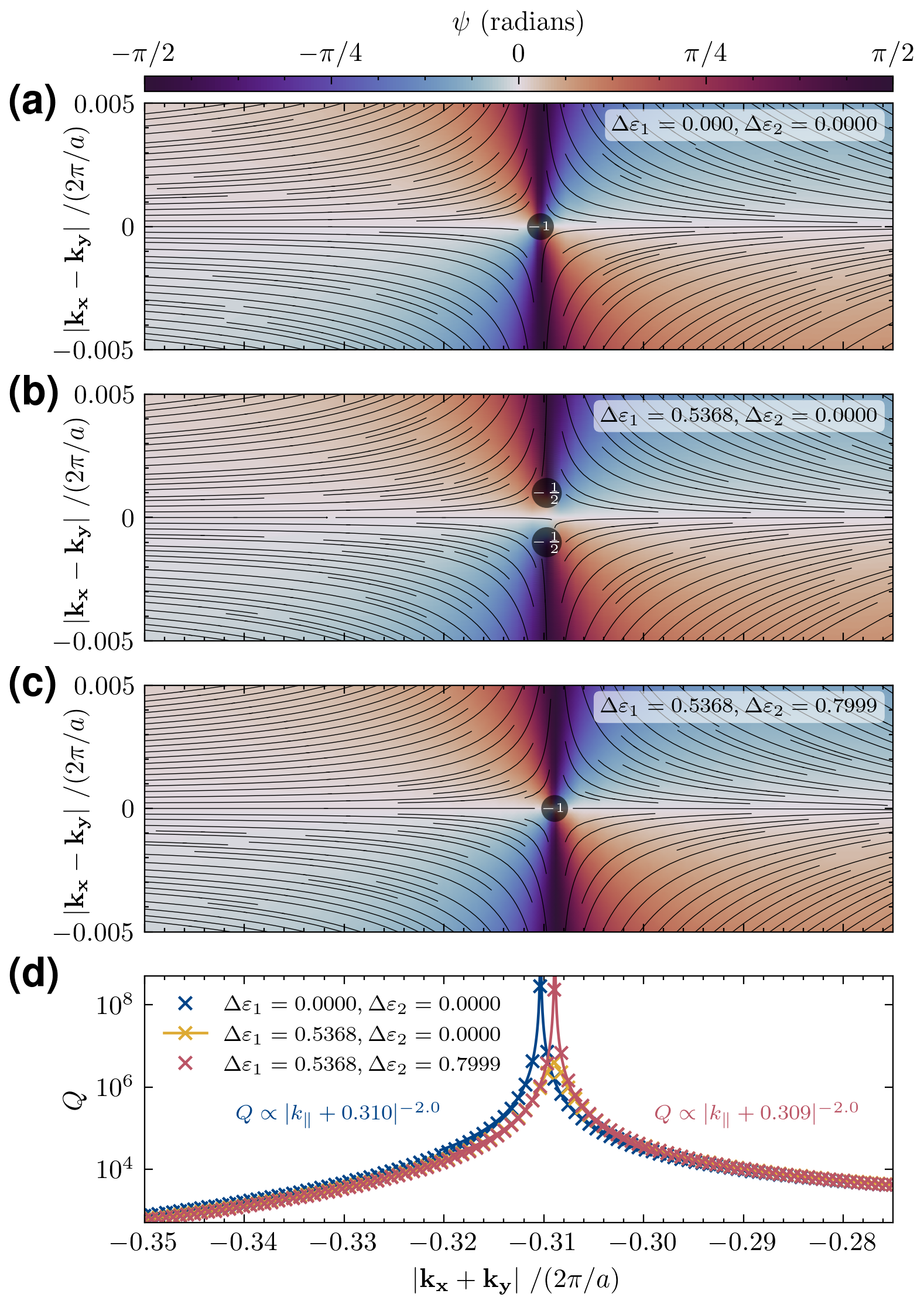} 

	\caption{Far-field polarization and $Q$--$\mathbf{k}$ relation of the tracked BIC mode under different perturbation parameters.
		(a)--(c) Far-field polarization of the tracked BIC mode for
		(a) $(\Delta \varepsilon_1, \Delta \varepsilon_2)=(0,0)$,
		(b) $(0.5368,0)$, and
		(c) $(0.5368,0.7999)$.
		The color represents the angle $\psi$ of the polarization vector, and the streamlines represent the polarization director field. The topological charge of the polarization vortex is marked in the figure.
		(d) The $Q$-$k$ relation of the tracked BIC mode with different perturbation parameters. BIC modes are fitted by $Q=\alpha \Delta k^{-n}$, where $\alpha$ and $n$ are fitting parameters. Fitted results are shown in the figure.
	}
	\label{fig:2-recovery} 
\end{figure}

The tracked mode exhibits the standard signatures of a tunable BIC. As shown in Fig.~\ref{fig:2-recovery}(d), the $Q$ factor diverges as $Q\propto \Delta k^{-2}$ near the singular $\mathbf{k}$ point, where $\Delta k$ is the distance from the BIC $\mathbf{k}$ point. The far-field polarization (Fig.~\ref{fig:2-recovery}(a)) also reveals a polarization vortex with a topological charge of $-1$.

We then consider the $\sigma_z$-broken condition with permittivity of cladding $\varepsilon_{\mathrm{up}}$ and $\varepsilon_{\mathrm{down}}$, which is a common configuration for integrated photonic devices. This causes the tunable BIC to degrade into a quasi-BIC with a finite $Q$ factor (Fig.~\ref{fig:1-band}(e)).
To quantify perturbation $\hat{V}_a$ introduced by substrate, we define
\begin{equation}
	\Delta \varepsilon_1=\frac{1}{2}(\varepsilon_{\mathrm{down}}-\varepsilon_{\mathrm{up}})\label{eq:delta-eps1}
\end{equation}
to describe the intensity $\lambda_a$. In this case, $\Delta \varepsilon_1=0.5368$.

The $Q$ factor of the tracked BIC decreases from infinity to about $10^6$ (Fig.~\ref{fig:2-recovery}(d)). The far-field polarization is shown in Fig.~\ref{fig:2-recovery}(b), which clearly shows the splitting of the topological charge $-1$ into two half-charges $-1/2$ located on either side of the $\Gamma$-M direction.

We design another odd-parity perturbation $\hat{V}_b$ by simply applying a bilayer photonic crystal slab, which has different permittivity distribution along the $z$-direction. It should be noted that the bilayer photonic crystal slab is not the only way to implement odd-parity perturbation. Alternative strategies, such as depositing an extra layer onto the slab surface or tapering the cylindrical air holes into truncated cones, can also be used to implement the secondary perturbation. A comprehensive discussion of various odd-parity perturbation schemes is provided in the Supplemental Material (\cite{SI}, Sec.~S3).

We define $\Delta \varepsilon_2$ to quantify the intensity $\lambda_b$ of the perturbation $\hat{V}_b$, which can be expressed as
\begin{equation}
	\Delta \varepsilon_2 = \frac{1}{2}(\varepsilon_{\mathrm{l1}}-\varepsilon_{\mathrm{l2}}),
\end{equation}
where $\varepsilon_{\mathrm{l1}}$ and $\varepsilon_{\mathrm{l2}}$ are the permittivity of the top and bottom layer materials of the photonic crystal, respectively.

By tuning $\Delta \varepsilon_2$, we can recover tunable BICs with nonzero $\Delta \varepsilon_1$. For example, when applying $\Delta \varepsilon_1=0.5368$, the recovery condition is reached at $\Delta \varepsilon_2=0.7999$.
The tracked BIC again shows $Q\propto \Delta k^{-2}$ and diverges at $k_{\mathrm{BIC}}$ (Fig.~\ref{fig:1-band}(f)). The separation of the two half-charges in the far-field polarization is also eliminated, and they merge back to a full charge of -1 (Fig.~\ref{fig:2-recovery}(c)).

\begin{figure}
	\centering
	\includegraphics[width=0.48\textwidth]{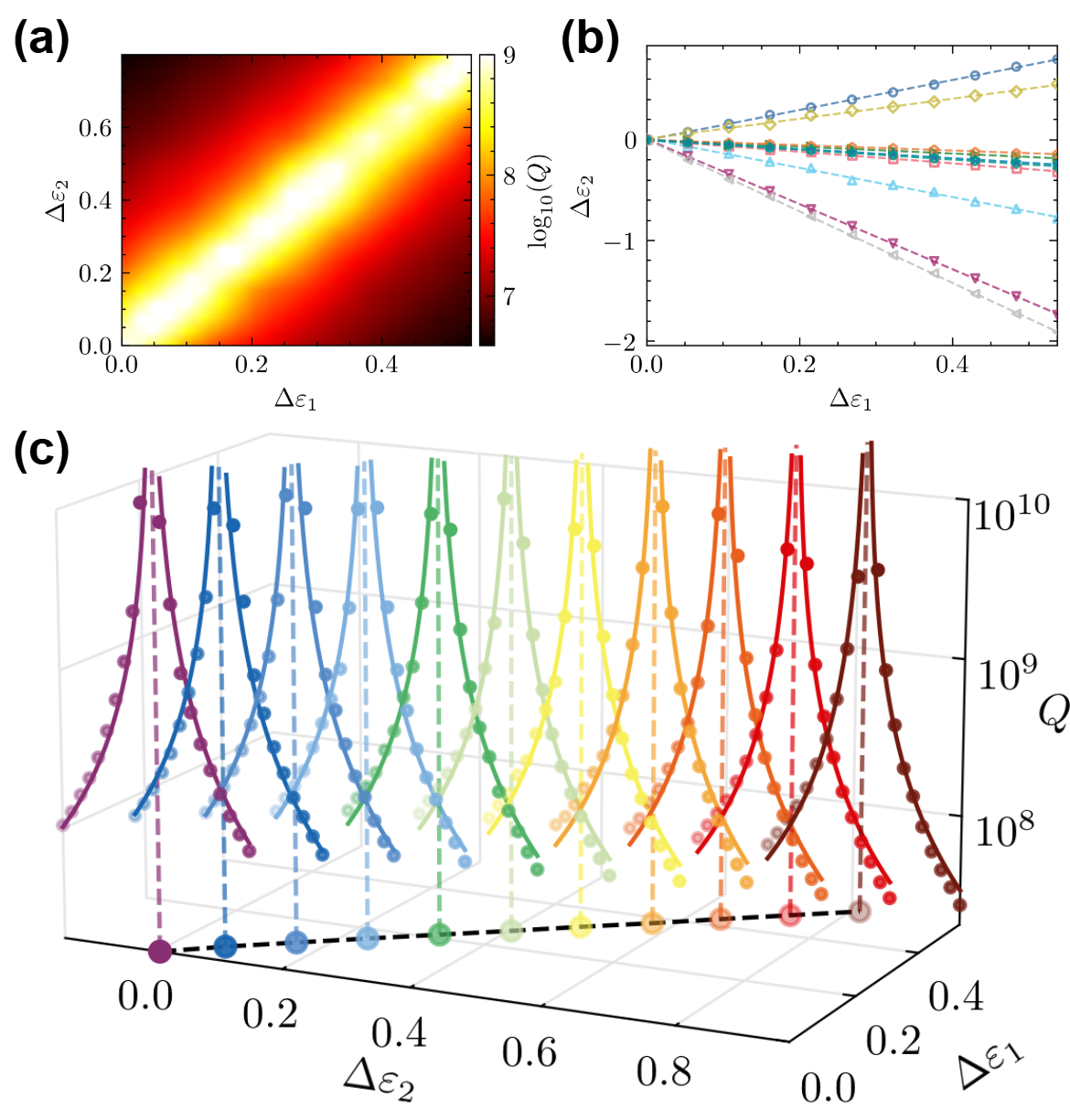} 

	\caption{Parameter dependence of the $Q$ factor for the tracked BIC mode.
		(a) $Q$ factor mapping of the tracked BIC mode or degraded quasi-BIC mode in the $(\Delta \varepsilon_1, \Delta \varepsilon_2)$-space. The color represents the $\log_{10}(Q)$ factor.
		(b) $\Delta \varepsilon_1$-$\Delta \varepsilon_2$ relation of the recovery lines for different tunable BIC modes. Each line corresponds to a tunable BIC mode.
		(c) $Q$ factor relation of the tracked BIC mode with $\Delta \varepsilon_2$ at a fixed $k_{\mathrm{BIC}}$ point. The dots are calculated results and the solid line is fitted by $Q=\alpha |\Delta \varepsilon_2 - \Delta \varepsilon_{2,\mathrm{rec}}|^{-2}$, where $\alpha$ is the fitting parameter and $\Delta \varepsilon_{2,\mathrm{rec}}$ is the fitted recovery value of $\Delta \varepsilon_2$.
	}
	\label{fig:3-recovery} 
\end{figure}

The recovery is not an isolated point in the parameter space. We calculate the $Q$ factor of BIC (or quasi-BIC) in the $(\Delta \varepsilon_1, \Delta \varepsilon_2)$-space and plot the result in Fig.~\ref{fig:3-recovery}(a). The recovery parameters are clearly linearly related, which is consistent with the theoretical analysis Eq.~\ref{eq:det-d-ab}.
Accordingly, the recovery points satisfy an approximately linear relation, $\lambda_b=c\,\lambda_a$, where $c$ is determined by the modal overlap integrals.

Also, the recovery is not limited to the tracked BIC. We can recover any tunable BICs by tuning the two perturbation parameters. As shown in Fig.~\ref{fig:3-recovery}(b), recovery lines exist for different tunable BIC modes, each governed by a distinct linear scaling relation.

The relation between $Q$ factor and $\Delta \varepsilon_2$ at a fixed $\Delta \varepsilon_1$ and $\mathbf{k}$ point is shown in Fig.~\ref{fig:3-recovery}(c). The $\mathbf{k}$ point is chosen to be the same as the recovered tracked BIC. The $Q$ factor diverges at the recovery point and shows $Q \propto 1/\gamma \propto |\Delta \varepsilon_2 - \Delta \varepsilon_{2,\mathrm{rec}}|^{-2}$, where $\Delta \varepsilon_{2,\mathrm{rec}}$ is the recovery parameter of $\Delta \varepsilon_2$. This provides strong evidence that the degraded quasi-BIC is indeed restored to an exact BIC~\cite{koshelev_Asymmetric_2018}.

This recovery mechanism is particularly valuable because it operates in substrate-supported photonic crystals, making tunable BICs directly relevant to integrated photonic devices. A more stringent test of the mechanism is provided by the merging-BIC configuration, which arises from the merging of tunable BICs with a symmetry-protected BIC~\cite{jin_Topologically_2019}. Such configurations are known to support robust ultrahigh-$Q$ resonances over a broadened momentum-space region compared with conventional photonic crystal designs. At the same time, they are highly vulnerable to substrate-induced symmetry breaking: once the tunable BICs involved in the merging process degrade into quasi-BICs, the merging-BIC configuration is destroyed and reduces to an ordinary symmetry-protected BIC.

Figures~\ref{fig:4-merging}(a) and \ref{fig:4-merging}(b) show the near merging configuration (see Supplemental Material~\cite{SI}, Sec.~S1 for details) under different perturbation conditions. In the mirror-symmetric structure, the tunable BICs occur at $\log_{10}(|k_{\parallel}|/(2\pi/a))=-1.47$, and the corresponding scaling follows
\begin{equation}
	Q\propto k_{\parallel}^{-2}(k_{\parallel}-k_{\mathrm{BIC}})^{-2}(k_{\parallel}+k_{\mathrm{BIC}})^{-2}.
\end{equation}
When the substrate perturbation is introduced with $\Delta \varepsilon_1=0.5368$, the tunable BICs degrade into quasi-BICs, and the scaling collapses to the ordinary symmetry-protected form $Q\propto k_{\parallel}^{-2}$.
After tuning the secondary perturbation to $\Delta \varepsilon_2=0.0609$ in the near merging configuration, the tunable BICs reappear and the characteristic high-$Q$ scaling is correspondingly restored.

The results for the exact merging-BIC configuration (see Supplemental Material~\cite{SI}, Sec.~S1 for details) are shown in Figs.~\ref{fig:4-merging}(c) and \ref{fig:4-merging}(d). The merging-BIC is restored using the same recovery parameters as those required for the near merging configuration, and the recovered configuration exhibits the characteristic scaling $Q\propto k_{\parallel}^{-6}$.
This demonstrates that the proposed recovery mechanism is capable of reconstructing not only isolated tunable BICs, but also the higher-order topological configurations built from them.

To further demonstrate this generality, we consider two additional implementations of the secondary perturbation, namely a tapered-hole structure and an extra surface layer structure.
The unperturbed configurations are the same as those in the bilayer design. The tapered-hole structure is implemented by tapering the cylindrical air holes into truncated cones with fixed average radius and varying taper angle $\theta$. The extra surface layer structure is implemented by adding layers with thickness $t=0.1a$ on the top and bottom of the photonic crystal slab, with unperturbed permittivity $\varepsilon_{\mathrm{el0}}=\varepsilon_{\sigma_z}$ and varying permittivity difference $\Delta \varepsilon_{\mathrm{el}}$ between the two layers.
As shown in Figs.~\ref{fig:4-merging}(e) and \ref{fig:4-merging}(f), both implementations exhibit an approximately linear relation between the substrate-induced asymmetry and the corresponding recovery parameter, similar to that obtained in the bilayer design. The different slopes reflect the different spatial profiles and modal-overlap coefficients of the perturbations. Details of these structures are provided in the Supplemental Material~\cite{SI}, Sec. S3.

\begin{figure}
	\centering
	\includegraphics[width=0.48\textwidth]{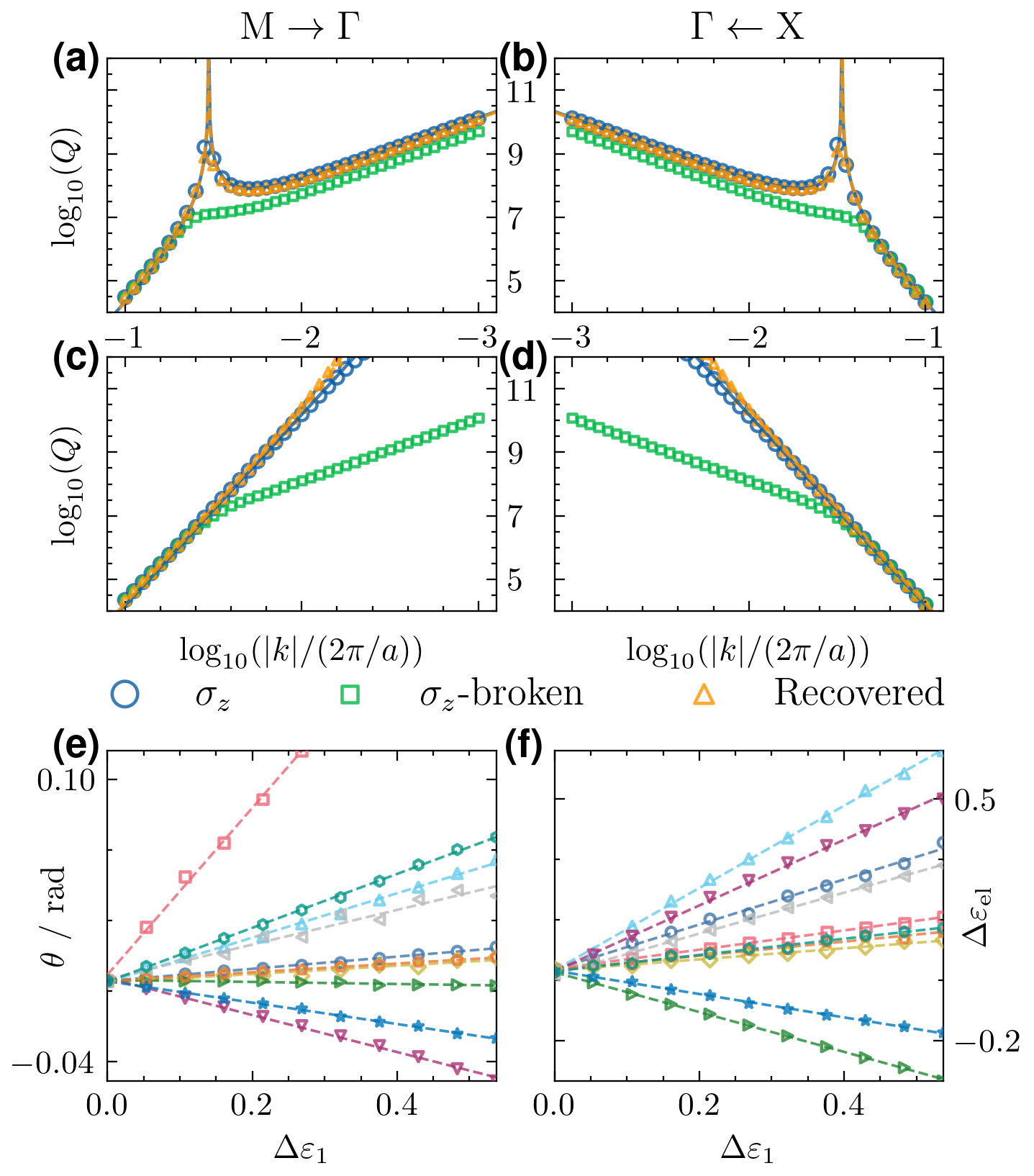} 

	\caption{$\log_{10}(Q)$--$\log_{10}(|k_{\parallel}|)$ relation of (near) merging-BIC configurations and the linear recovery relations for different secondary perturbation implementations.
		(a),(b) Near-merging configuration. (c),(d) Exact merging-BIC configuration. The perturbation parameters are $(\Delta \varepsilon_1, \Delta \varepsilon_2)=(0,0)$ for the mirror-symmetric case ($\sigma_z$), $(0.5368,0)$ for the substrate-broken case ($\sigma_z$-broken), and $(0.5368,0.0609)$ for the recovered case (Recovered). (a),(c) Band structures along the $\mathrm{M}\rightarrow \Gamma$ direction.
		(b),(d) Band structures along the $\Gamma\leftarrow \mathrm{X}$ direction.
		The $\sigma_z$ and recovered lines are fitted by $Q=\alpha k_{\parallel}^{-2}(k_{\parallel}+k_{\mathrm{BIC}})^{-2}(k_{\parallel}-k_{\mathrm{BIC}})^{-2}$, where $\alpha$ is the fitting parameter and $k_{\mathrm{BIC}}$ is the fitted tunable BIC $\mathbf{k}$ point.
		(e) and (f) show the linear recovery relations obtained using tapered air holes and an extra surface layer as the secondary perturbation, respectively.
	}
	\label{fig:4-merging} 
\end{figure}


In conclusion, we have shown that tunable BICs destroyed by substrate-induced mirror-symmetry breaking can be recovered by introducing a secondary odd-parity perturbation. The mechanism is captured by a simple linear cancellation condition in the radiation matrix and verified numerically. The recovered states regain both the topological polarization vortex and the characteristic $Q\propto \Delta k^{-2}$ scaling, despite the presence of strong environmental asymmetry. More broadly, our results establish dual-asymmetry compensation as a general strategy for preserving tunable BICs in practical photonic architectures, opening new avenues for the deployment of high-$Q$ topological states in integrated nanophotonic devices.

\begin{acknowledgments}
	This work is supported by the National Key Research and Development Program of China (2024YFE0204700), NSFC grant (62504195), Guangdong Basic and Applied Basic Research Foundation (2026A1515011332), Guangdong Province Project (2024QN11C061), University Development Fund of The Chinese University of Hong Kong (Shenzhen) under the grant UDF01003700 and UDF01003324, and SSE Research Fund (G10120250477).
\end{acknowledgments}

\bibliography{reference}

\end{document}



\begin{center}
	{\LARGE\bfseries Supplemental Material}\\[1em]
\end{center}

\title{Recovery of tunable bound states in the continuum}




\author{Hai~Huang}
\author{Huiming~Zhang}
\author{Biye~Xie}
\author{Wengang~Bi}
\email[]{waynebi@cuhk.edu.cn}
\author{Daquan~Zhang}
\email[]{zhangdaquan@cuhk.edu.cn}
\affiliation{%
	School of Science and Engineering, The Chinese University of Hong Kong (Shenzhen), Longgang, Shenzhen, Guangdong, 518172, P.R. China.
}%



\maketitle


\section{Numerical simulation methods}

All numerical simulations in this work are performed by solving the source-free Maxwell eigenvalue problem for a periodic photonic crystal (PhC) under Bloch boundary conditions using the finite-element method.
The lattice is square, with lattice constant $a=680~\mathrm{nm}$, chosen to place the tracked BICs near $\lambda_0=1550~\mathrm{nm}$.
The photonic crystal slab has thickness $t=0.5a$ and the air holes have radius $r=0.2a$. For the $\sigma_z$-broken system, the relative permittivity is $\varepsilon_{\mathrm{up}}=1$ for the top cladding and $\varepsilon_{\mathrm{down}}=2.0736=1.44^2$ for the bottom cladding~\cite{malitson_Interspecimen_1965}. The relative permittivity of the PhC slab is $\varepsilon_{\mathrm{PhC}}=14.2129=3.77^2$~\cite{karaman_Decoupling_2025}.

For near merging-BIC configuration, thickness of photonic crystal slab is $t=1.132a$ and the radius of air hole is $r=0.292t$. For exact merging-BIC configuration, thickness of photonic crystal slab is $t=1.128a$ and the radius of air hole is $r=0.292t$.

The computational domain is truncated by perfectly matched layers (PMLs) in the $z$ direction. The PML thickness is $0.5\lambda_0$, and the cladding regions have a thickness of $2\lambda_0$ to ensure negligible coupling between the resonant modes and the PMLs.

For band-structure calculations, the mesh is refined to ensure that modes with $Q>10^{10}$ could be reliably identified, and the step size in the wave-vector sweep is $10^{-3}\times\pi/a$. For far-field polarization extraction, the sweeping step size is $10^{-3} \times 2\pi/a$ in $\Gamma-\mathrm{M}$ direction and $\sqrt{2}\times 10^{-4} \times 2\pi/a$ in orthogonal directions. For merging-BIC calculations, the mesh is further refined to ensure $Q>10^{12}$ can be reliably identified.

All fitting results are obtained using the SciPy library in Python.

The far-field polarization vector is extracted from the upward radiation channel. Its angle is denoted by $\psi$, and the polarization topology is analyzed by tracking the winding of the far-field polarization vector around the singular point. The associated topological charge is calculated from
\begin{equation}
	q=\frac{1}{2\pi}\oint_C \nabla_{\mathbf{k}} \psi(\mathbf{k})\cdot d\mathbf{k},
\end{equation}
where $C$ is a closed loop in momentum space enclosing the singularity. In the figures, the streamlines correspond to the polarization director field.

\section{Details of derivations}

\subsection{Friedrich-Wintgen BICs and singularity of the radiation matrix}

Friedrich-Wintgen (FW) BICs arise from destructive interference between different resonant modes coupled to the same radiation continuum. In temporal coupled-mode theory, the two-mode non-Hermitian Hamiltonian can be written as
\begin{equation}
	H=\bm{\Omega}-i\bm{\Gamma}
	=\begin{pmatrix}
		\omega_1 & \kappa   \\
		\kappa   & \omega_2
	\end{pmatrix}
	-i\begin{pmatrix}
		\gamma_1   & \gamma_0 \\
		\gamma_0^* & \gamma_2
	\end{pmatrix},
	\label{eq:S_Hamiltonian}
\end{equation}
where $\omega_1$ and $\omega_2$ are the modal frequencies, $\kappa$ is the near-field coupling coefficient, $\gamma_1$ and $\gamma_2$ are the radiative decay rates, and $\gamma_0$ is the far-field coupling coefficient between the two modes.

Let $|\psi_{\mathrm{BIC}}\rangle$ be the eigenstate associated with a BIC. By definition, its eigenvalue must be real:
\begin{equation}
	H|\psi_{\mathrm{BIC}}\rangle=\omega_{\mathrm{BIC}}|\psi_{\mathrm{BIC}}\rangle,
\end{equation}
where $\omega_{\mathrm{BIC}}\in\mathbb{R}$. Left-multiplying by $\langle\psi_{\mathrm{BIC}}|$ gives
\begin{equation}
	\langle\psi_{\mathrm{BIC}}|\bm{\Omega}|\psi_{\mathrm{BIC}}\rangle
	-i\langle\psi_{\mathrm{BIC}}|\bm{\Gamma}|\psi_{\mathrm{BIC}}\rangle
	=\omega_{\mathrm{BIC}}\langle\psi_{\mathrm{BIC}}|\psi_{\mathrm{BIC}}\rangle.
	\label{eq:S_BIC_expectation}
\end{equation}
Since $\bm{\Omega}$ is Hermitian and $\omega_{\mathrm{BIC}}$ is real, the imaginary part of Eq.~\ref{eq:S_BIC_expectation} must vanish, which requires
\begin{equation}
	\langle\psi_{\mathrm{BIC}}|\bm{\Gamma}|\psi_{\mathrm{BIC}}\rangle=0.
\end{equation}
Because $\bm{\Gamma}$ is defined to be positive semidefinite (no gain in the system), this implies that $|\psi_{\mathrm{BIC}}\rangle$ belongs to the null space of $\bm{\Gamma}$, so $\bm{\Gamma}$ must be singular:
\begin{equation}
	\det(\bm{\Gamma})=0.
	\label{eq:S_detGamma0}
\end{equation}

The radiation-coupling matrix for a two-mode system can be expressed in terms of the coupling amplitudes as
\begin{equation}
	\bm{D}=\begin{pmatrix}
		d_1^+ & d_2^+ \\
		d_1^- & d_2^-
	\end{pmatrix},
	\label{eq:S_Dmatrix}
\end{equation}
where $d_i^+$ and $d_i^-$ are the coupling amplitudes of the $i$-th mode to the upward and downward radiation channels, respectively. Energy conservation requires
\begin{equation}
	\bm{D}^\dagger \bm{D}=2\bm{\Gamma}.
	\label{eq:S_DDgamma}
\end{equation}
Therefore,
\begin{equation}
	\det(\bm{\Gamma})=\frac{1}{4}\det(\bm{D}^\dagger\bm{D})=\frac{1}{4}|\det(\bm{D})|^2.
\end{equation}
Thus the singularity of $\bm{\Gamma}$ is equivalent to the singularity of $\bm{D}$:
\begin{equation}
	\det(\bm{\Gamma})=0\Longleftrightarrow \det(\bm{D})=0
	\Longleftrightarrow d_1^+d_2^- - d_1^-d_2^+=0.
	\label{eq:S_detD_condition}
\end{equation}
Eq.~\ref{eq:S_detD_condition} is the \textbf{necessary} condition for the existence of a BIC.

Once Eq.~\ref{eq:S_detD_condition} is satisfied, the remaining condition for a two-mode FW BIC reduces to
\begin{equation}
	\kappa(\gamma_1-\gamma_2)=\pm \sqrt{\gamma_1\gamma_2}(\omega_1-\omega_2).
	\label{eq:S_FW_condition}
\end{equation}
In $\sigma_z$-symmetric systems, Eq.~\ref{eq:S_detD_condition} is automatically satisfied. We can adjust $(k_x,k_y)$ to satisfy Eq.~\ref{eq:S_FW_condition}, which is an additional constraint that must be satisfied to ensure the existence of a BIC.

\subsection{Derivation of the symmetry relations}

The radiation coupling amplitudes of $i$-th mode are defined as the overlap between the resonant mode and the open $\pm z$ radiation channel. It can be written as
\begin{equation}
	d_i^\pm \propto \langle \Phi^\pm | \hat{V} | \Psi_i \rangle,
	\label{eq:S_general_d}
\end{equation}
where $|\Psi_i\rangle$ is the resonant mode, $|\Phi^\pm\rangle$ are the normalized outgoing radiation channels propagating toward $\pm z$, and $\hat{V}$ is the radiation-coupling operator (details of the operator can be found in Refs.~\cite{jin_Topologically_2019,yin_Topological_2023}).

Under the mirror operator $\hat{\sigma}_z$, we define the channel transformation by
\begin{equation}
	\hat{\sigma}_z |\Phi^-\rangle = \chi |\Phi^+\rangle,
	\label{eq:S_channel_transform}
\end{equation}
where $\chi$ is a unit-modulus phase factor $\chi=e^{i\Delta\theta}$ determined by the channel phase difference. By a suitable choice of channel gauge, it can be set to $\chi=1$. This choice is a convention and does not affect any conclusion.

For the unperturbed mirror-symmetric system, the unperturbed coupling operator $\hat{V}_0$ is invariant under $\hat{\sigma}_z$. If the unperturbed resonant mode has definite parity $p_0=\pm1$, then
\begin{equation}
	\hat{\sigma}_z |\Psi_i^{(0)}\rangle = p_0 |\Psi_i^{(0)}\rangle.\label{eq:state_d0}
\end{equation}
Using Eq.~\ref{eq:S_general_d}, we obtain
\begin{equation}
	d_i^{-(0)} \propto \langle \Phi^-|\hat{V}_0|\Psi_i^{(0)}\rangle.
\end{equation}
Substituting Eq.~\ref{eq:S_channel_transform} and using mirror symmetry,
\begin{equation}
	\begin{aligned}
		d_i^{-(0)}
		 & \propto \langle \Phi^-|\hat V_0|\Psi_i^{(0)}\rangle                \\
		 & = \chi\langle \Phi^+|\hat\sigma_z^{-1}\hat V_0|\Psi_i^{(0)}\rangle \\
		 & = \chi\langle \Phi^+|\hat V_0\hat\sigma_z^{-1}|\Psi_i^{(0)}\rangle \\
		 & = \chi p_0\langle \Phi^+|\hat V_0|\Psi_i^{(0)}\rangle .
	\end{aligned}
\end{equation}
Therefore,
\begin{equation}
	d_i^{-(0)}=\chi p_0 d_i^{+(0)}.
	\label{eq:S_d0_general}
\end{equation}
Using the gauge choice $\chi=1$, this reduces to
\begin{equation}
	d_i^{-(0)}=p_0 d_i^{+(0)}.
	\label{eq:S_d0_maintext}
\end{equation}
This is the relation used in the main text.

Then considering an odd-parity perturbation $\hat{V}_a$ with intensity $\lambda_a$ and parity $p_a=-1$, we expand the radiation-coupling operator and the resonant mode as
\begin{equation}
	\hat V = \hat V_0 + \lambda_a \hat V_a,
	\qquad
	|\Psi_i\rangle = |\Psi_i^{(0)}\rangle + \lambda_a |\Psi_i^{(1)}\rangle + \mathcal{O}(\lambda_a^2).
\end{equation}
Recalling Eq.~\ref{eq:S_general_d}, and substituting the perturbative expansions, we obtain
\begin{equation}
	d_i^\pm = d_i^{\pm(0)} + \lambda_a c_{ia}^\pm + \mathcal{O}(\lambda_a^2),
\end{equation}
where
\begin{equation}
	d_i^{\pm(0)} \propto \langle \Phi^\pm | \hat V_0 | \Psi_i^{(0)} \rangle,
\end{equation}
and
\begin{equation}
	c_{ia}^\pm \propto
	\langle \Phi^\pm | \hat V_a | \Psi_i^{(0)} \rangle
	+
	\langle \Phi^\pm | \hat V_0 | \Psi_i^{(1)} \rangle.
	\label{eq:S_cia_definition}
\end{equation}
The first-order correction to the mode can be written as
\begin{equation}
	|\Psi_i^{(1)}\rangle
	=
	\sum_{m\neq i}
	\frac{\langle \Psi_m^{(0)}|\hat V_a|\Psi_i^{(0)}\rangle}
	{\omega_i^{(0)}-\omega_m^{(0)}}
	|\Psi_m^{(0)}\rangle.
\end{equation}
For an unperturbed mode with parity $p_0$ and an odd perturbation with parity $p_a=-1$, only states with parity $p_a p_0$ contribute to the above sum. Therefore,
\begin{equation}
	\hat{\sigma}_z |\Psi_i^{(1)}\rangle = p_a p_0 |\Psi_i^{(1)}\rangle.
	\label{eq:S_psi1_parity}
\end{equation}
We now define the first-order radiation source
\begin{equation}
	|S_{ia}^{(1)}\rangle
	\equiv
	\hat V_a|\Psi_i^{(0)}\rangle+\hat V_0|\Psi_i^{(1)}\rangle,
\end{equation}
so that
\begin{equation}
	c_{ia}^\pm \propto \langle \Phi^\pm | S_{ia}^{(1)} \rangle.
\end{equation}
Using the parity transformation of the operators and states in Eq.~\ref{eq:state_d0} and Eq.~\ref{eq:S_psi1_parity}, together with the mirror-symmetry properties of the operators,
\begin{equation}
	\hat{\sigma}_z \hat V_0 \hat{\sigma}_z^{-1}=\hat V_0,
	\qquad
	\hat{\sigma}_z \hat V_a \hat{\sigma}_z^{-1}=p_a \hat V_a,
	\label{eq:S_operator_parity}
\end{equation}
we can determine the parity of the first-order radiation source $|S_{ia}^{(1)}\rangle$.

For the first term, we have
\begin{equation}
	\begin{aligned}
		\hat{\sigma}_z \hat V_a |\Psi_i^{(0)}\rangle
		 & =
		\left(\hat{\sigma}_z \hat V_a \hat{\sigma}_z^{-1}\right)
		\left(\hat{\sigma}_z |\Psi_i^{(0)}\rangle\right) \\
		 & =
		p_a \hat V_a \, p_0 |\Psi_i^{(0)}\rangle         \\
		 & =
		p_a p_0 \hat V_a |\Psi_i^{(0)}\rangle.
	\end{aligned}
\end{equation}
For the second term, similarly,
\begin{equation}
	\begin{aligned}
		\hat{\sigma}_z \hat V_0 |\Psi_i^{(1)}\rangle
		 & =
		\left(\hat{\sigma}_z \hat V_0 \hat{\sigma}_z^{-1}\right)
		\left(\hat{\sigma}_z |\Psi_i^{(1)}\rangle\right) \\
		 & =
		\hat V_0 \, p_a p_0 |\Psi_i^{(1)}\rangle         \\
		 & =
		p_a p_0 \hat V_0 |\Psi_i^{(1)}\rangle.
	\end{aligned}
\end{equation}
Therefore, both terms in $|S_{ia}^{(1)}\rangle$ have the same parity $p_a p_0$, and thus the first-order radiation source satisfies
\begin{equation}
	\hat{\sigma}_z |S_{ia}^{(1)}\rangle = p_a p_0 |S_{ia}^{(1)}\rangle.
	\label{eq:S_source_parity}
\end{equation}
Now consider the first-order correction to the downward radiation amplitude:
\begin{equation}
	c_{ia}^- \propto \langle \Phi^- | S_{ia}^{(1)} \rangle.
\end{equation}
Using Eq.~\ref{eq:S_channel_transform}, we rewrite
\begin{equation}
	\begin{aligned}
		c_{ia}^-
		 & \propto \langle \Phi^- | S_{ia}^{(1)} \rangle                 \\
		 & = \chi \langle \Phi^+|\hat{\sigma}_z^{-1}|S_{ia}^{(1)}\rangle \\
		 & = \chi p_a p_0 \langle \Phi^+|S_{ia}^{(1)}\rangle.
	\end{aligned}
\end{equation}
Hence, we obtain
\begin{equation}
	c_{ia}^-=\chi p_a p_0 c_{ia}^+.
	\label{eq:S_cia_general}
\end{equation}
Under the gauge choice $\chi=1$, this reduces to
\begin{equation}
	c_{ia}^- = p_a p_0 c_{ia}^+.
	\label{eq:S_cia_maintext}
\end{equation}
For the odd-parity perturbation considered here, $p_a=-1$, and therefore
\begin{equation}
	c_{ia}^- = -p_0 c_{ia}^+.
	\label{eq:S_cia_odd}
\end{equation}
This is the relation used in the main text.

\subsection{Recovery under second odd-parity perturbation}

The symmetry relations derived above can be straightforwardly extended to the case of two odd-parity perturbations. Let $\hat{V}_a$ and $\hat{V}_b$ be two odd-parity perturbations with strengths $\lambda_a$ and $\lambda_b$, respectively. To first order, the radiation coupling amplitudes can be written as
\begin{equation}
	d_i^\pm = d_i^{\pm(0)} + c_{ia}^\pm \lambda_a + c_{ib}^\pm \lambda_b.
\end{equation}
Using the symmetry relations in Eq.~\ref{eq:S_d0_maintext} and Eq.~\ref{eq:S_cia_odd}, we can rewrite the determinant of the radiation-coupling matrix as
\begin{equation}
	\begin{aligned}
		\det(\bm{D})
		= & 2p_0\bigl(d_2^{+(0)}c_{1a}^+ - d_1^{+(0)}c_{2a}^+\bigr)\lambda_a  \\
		+ & 2p_0\bigl(d_2^{+(0)}c_{1b}^+ - d_1^{+(0)}c_{2b}^+\bigr)\lambda_b.
	\end{aligned}
\end{equation}
The singularity condition $\det(\bm{D})=0$ can be satisfied by tuning $\lambda_b$ to
\begin{equation}
	\lambda_b = -\frac{d_2^{+(0)}c_{1a}^+ - d_1^{+(0)}c_{2a}^+}{d_2^{+(0)}c_{1b}^+ - d_1^{+(0)}c_{2b}^+}\lambda_a.
\end{equation}

\section{Alternative odd-parity perturbations}

In this section, we discuss different odd-parity and quasi-odd-parity perturbations $\hat{V}_x$ with effective parity $p_x=-1$ and strength $\lambda_x$.

\subsection{Substrate}
We redefine the claddings of the symmetric reference system as in Eq.~6 of the main text, thus the unperturbed system permittivity is
\begin{equation}
	\varepsilon_0(\mathbf{r}) = \begin{cases}
		\frac{\varepsilon_{\mathrm{up}}+\varepsilon_{\mathrm{down}}}{2} & z>\frac{t}{2}       \\
		\varepsilon_{\mathrm{PhC}}(\mathbf{r})                          & |z|\leq \frac{t}{2} \\
		\frac{\varepsilon_{\mathrm{up}}+\varepsilon_{\mathrm{down}}}{2} & z<-\frac{t}{2}
	\end{cases}.
\end{equation}
The substrate-induced perturbation can be written as
\begin{equation}
	{\Delta \varepsilon}_{\mathrm{sub}}(\mathbf{r}) = \begin{cases}
		\frac{\varepsilon_{\mathrm{up}}-\varepsilon_{\mathrm{down}}}{2} & z>\frac{t}{2}       \\
		0                                                               & |z|\leq \frac{t}{2} \\
		\frac{\varepsilon_{\mathrm{down}}-\varepsilon_{\mathrm{up}}}{2} & z<-\frac{t}{2}
	\end{cases}.
\end{equation}
It is clear that ${\Delta \varepsilon}_{\mathrm{sub}}$ is an odd-parity perturbation; when the intensity $\lambda_x$ is 1, $\hat{V}_x$ corresponds to the actual substrate loading.

\subsection{Bilayer photonic crystal}
The dual-layer photonic crystal slab which is used to implement the second odd-parity perturbation in the main text can also be regarded as an odd-parity perturbation to a single-layer photonic crystal slab. The unperturbed system is the single-layer photonic crystal slab with uniform permittivity $\varepsilon_l=(\varepsilon_{\mathrm{l1}}+\varepsilon_{\mathrm{l2}})/2$,
\begin{equation}
	\varepsilon_0(\mathbf{r}) = \begin{cases}
		\frac{\varepsilon_{\mathrm{up}}+\varepsilon_{\mathrm{down}}}{2} & |z|>\frac{t}{2}                        \\
		\varepsilon_l                                                   & |z|\leq \frac{t}{2},\; x^2+y^2\geq r^2 \\
		\varepsilon_{\mathrm{hole}}                                     & |z|\leq \frac{t}{2},\; x^2+y^2<r^2
	\end{cases}.
\end{equation}
The bilayer slab can be regarded as an odd-parity perturbation to the single-layer slab, with the perturbation profile
\begin{equation}
	\Delta\varepsilon_{\mathrm{bilayer}}(\mathbf{r}) = \begin{cases}
		0                                                             & |z|>\frac{t}{2}                         \\
		\frac{\varepsilon_{\mathrm{l1}}-\varepsilon_{\mathrm{l2}}}{2} & 0<z\leq \frac{t}{2},\; x^2+y^2\geq r^2  \\
		\frac{\varepsilon_{\mathrm{l2}}-\varepsilon_{\mathrm{l1}}}{2} & -\frac{t}{2}<z\leq 0,\; x^2+y^2\geq r^2 \\
		0                                                             & |z|\leq \frac{t}{2},\; x^2+y^2<r^2
	\end{cases}.
\end{equation}
This profile is therefore odd with respect to $z$.

\subsection{Graded-index photonic crystal slab}

The bilayer photonic crystal slab is a simple discrete realization of an odd-parity perturbation. The same perturbation can also be implemented by a graded-index slab, in which the permittivity varies continuously along the $z$ direction.

We write the graded-index perturbation as
\begin{equation}
	\Delta\varepsilon_{\mathrm{g}}(\mathbf{r}) =
	\begin{cases}
		f(x,y)g(z), & |z|\leq t/2, \\
		0,          & |z|>t/2,
	\end{cases}
\end{equation}
where $f(x,y)$ describes the in-plane pattern and $g(z)$ describes the out-of-plane index gradient. If
\begin{equation}
	g(-z)=-g(z),
\end{equation}
then
\begin{equation}
	\Delta\varepsilon_{\mathrm{g}}(x,y,-z)
	=
	-\Delta\varepsilon_{\mathrm{g}}(x,y,z),
\end{equation}
so the graded-index profile is an odd-parity perturbation with respect to $\sigma_z$.

For example, a linear index gradient,
\begin{equation}
	g(z)=\frac{2z}{t},
	\qquad |z|\leq t/2,
\end{equation}
satisfies this condition. The bilayer model used in the main text can be regarded as the piecewise-constant limit of this graded-index perturbation, corresponding to $g(z)=\mathrm{sgn}(z)$ inside the slab.

\subsection{Tapered-hole}

The tapered-hole implementation can also recover tunable BICs, as reported in Ref.~\cite{xu_Tapered_2024}. However, strictly speaking, the exact tapered-hole profile does not satisfy the pure odd-parity condition adopted in our analytical theory.

To see this, we redefine the unperturbed system as a single-layer photonic crystal slab with cylindrical holes of radius $R_{\mathrm{avg}}$, and treat the tapered hole as a geometric perturbation relative to this cylindrical reference. Let
\begin{equation}
	\rho=\sqrt{x^2+y^2},
	\qquad
	R(z)=R_{\mathrm{avg}}+\delta R(z),
\end{equation}
where for a linear taper,
\begin{equation}
	\delta R(z)=z\tan\theta.
\end{equation}
Then the exact dielectric perturbation associated with the tapered hole can be written as
\begin{equation}
	\Delta\varepsilon_{\mathrm{taper}}(\rho,z)
	=
	(\varepsilon_{\mathrm{hole}}-\varepsilon_{\mathrm{slab}})
	\left[
		\Theta(R(z)-\rho)-\Theta(R_{\mathrm{avg}}-\rho)
		\right],
	\qquad |z|\le \frac{t}{2},
	\label{eq:S_taper_exact}
\end{equation}
where $\Theta(\cdot)$ is the Heaviside step function, and $\Delta\varepsilon_{\mathrm{taper}}=0$ outside the slab.

Although the radius displacement $\delta R(z)$ is an odd function of $z$, the exact perturbation profile in Eq.~\ref{eq:S_taper_exact} is generally not strictly odd, namely
\begin{equation}
	\Delta\varepsilon_{\mathrm{taper}}(\rho,-z)
	\neq
	-\Delta\varepsilon_{\mathrm{taper}}(\rho,z).
\end{equation}
Therefore, the tapered-hole structure should not be regarded as an exact realization of the pure odd-parity perturbation.

Nevertheless, in the small-angle limit, the tapered-hole perturbation can be linearized around the cylindrical reference. Expanding Eq.~\ref{eq:S_taper_exact} to first order in the boundary displacement gives
\begin{equation}
	\Delta\varepsilon_{\mathrm{taper}}^{(1)}(\rho,z)
	\propto
	(\varepsilon_{\mathrm{hole}}-\varepsilon_{\mathrm{slab}})
	\,\delta(\rho-R_{\mathrm{avg}})\,\delta R(z),
	\label{eq:S_taper_first_order}
\end{equation}
where $\delta(\cdot)$ is the Dirac delta function. Since $\delta R(-z)=-\delta R(z)$, the leading-order perturbation satisfies
\begin{equation}
	\Delta\varepsilon_{\mathrm{taper}}^{(1)}(\rho,-z)
	=
	-\Delta\varepsilon_{\mathrm{taper}}^{(1)}(\rho,z).
\end{equation}
Hence, it can be regarded as an approximate or effective realization whose leading-order perturbation is odd-parity.

\subsection{Extra surface layer}

Related surface-layer-based recovery strategies have been reported in Refs.~\cite{bai_Recovery_2025,zhu_Observation_2026}. They considered an extra high-permittivity layer on top of the slab and an underetched slab, respectively.
Both implementations can be regarded as adding extra surface layers with permittivity $\varepsilon_{\mathrm{ex}}$ to the top and bottom surface of the slab. We redefine the unperturbed system as a single-layer photonic crystal slab with two extra surface layers on top and bottom, with the same thickness $d$,
\begin{equation}
	\varepsilon_0(\mathbf{r}) = \begin{cases}
		\varepsilon_{\mathrm{PhC}}(\mathbf{r})                          & |z|\leq \frac{t}{2}               \\
		\frac{\varepsilon_{\mathrm{ex}}+\varepsilon_{\mathrm{slab}}}{2} & \frac{t}{2}<|z|\leq \frac{t}{2}+d \\
		\frac{\varepsilon_{\mathrm{up}}+\varepsilon_{\mathrm{down}}}{2} & |z|>\frac{t}{2}+d
	\end{cases}.
\end{equation}
In this case, the extra surface layer can be regarded as an odd-parity perturbation to the single-layer slab, with the perturbation profile
\begin{equation}
	{\Delta \varepsilon}_{\mathrm{surface}}(\mathbf{r}) = \begin{cases}
		0                                                               & |z|\leq \frac{t}{2}               \\
		\frac{\varepsilon_{\mathrm{ex}}-\varepsilon_{\mathrm{slab}}}{2} & \frac{t}{2}<z\leq \frac{t}{2}+d   \\
		\frac{\varepsilon_{\mathrm{slab}}-\varepsilon_{\mathrm{ex}}}{2} & -\frac{t}{2}-d<z\leq -\frac{t}{2} \\
		0                                                               & |z|>\frac{t}{2}+d
	\end{cases}.
\end{equation}
${\Delta \varepsilon}_{\mathrm{surface}}$ is an odd-parity perturbation, and a very special case is when $\varepsilon_{\mathrm{ex}}=1$ and $\lambda_x=1$, the perturbation corresponds to the underetching method in Ref.~\cite{zhu_Observation_2026}. Also, when $\varepsilon_{\mathrm{ex}}=\varepsilon_{\mathrm{hole}}$ and $\lambda_x=1$, the perturbation corresponds to the extra layer method in Ref.~\cite{bai_Recovery_2025}.
By appropriately choosing $\varepsilon_{\mathrm{ex}}$ and $d$, the surface-layer method can be used to implement the second odd-parity perturbation for BIC recovery.

\section{Fabrication feasibility and combined tuning range}

The implementations discussed above can be connected to experimentally accessible fabrication routes. Bilayer and extra-layer perturbations can be realized through material growth, deposition, or post-growth treatment. Tapered holes can be introduced by anisotropic dry etching, such as reactive-ion etching, while graded-index perturbations can be implemented using semiconductor alloys with a continuously varying composition along the growth direction.

The recovery condition does not require the compensating perturbation to come from a single fabrication knob. If several odd-parity perturbations are present, their first-order radiation corrections add together, and the recovery point is determined by the combined odd-parity contribution. This allows, for example, a graded alloy profile to provide coarse tuning and a tapered-hole profile or extra dielectric layer to provide fine adjustment.

To compare different implementations on the same scale, Table~\ref{tab:S_feasible_range} summarizes representative tuning ranges and their relation to the recovery strengths used in this work.

\begin{table}[h]
	\caption{Representative tuning ranges for different odd-parity perturbation implementations and material platforms.}
	\label{tab:S_feasible_range}
	\begin{ruledtabular}
		\begin{tabular}{lll}
			Material/Method                                                 & Tuning range                                       & Ref.                                                \\
			\hline
			Bilayer photonic crystal                                        & $\Delta\varepsilon \in[-1.91, 0.80]$             & This work                                           \\
			Extra surface layer                                             & $\Delta\varepsilon \in[-0.32, 0.64]$ & This work                                           \\
			$\mathrm{Al}_x\mathrm{Ga}_{1-x}\mathrm{As}$ alloy @ 1.55 $\mu$m & $\varepsilon_1 \in [9.9214, 11.41]$                & \cite{papatryfonos_Refractive_2021} $x\in[0,0.452]$ \\
			$\mathrm{Al}_x\mathrm{Ga}_{1-x}\mathrm{As}$ alloy @ 0.9 $\mu$m  & $\varepsilon_1 \in [10.716, 12.478]$               & \cite{papatryfonos_Refractive_2021} $x\in[0,0.452]$ \\
			$\mathrm{Si}_x\mathrm{Ge}_{1-x}$ alloy @ 0.84 $\mu$m            & $\varepsilon_1 \in [13.476, 18.438]$               & \cite{jellison_Optical_1993} $x\in[0.2,0.98]$       \\
		\end{tabular}
	\end{ruledtabular}
\end{table}

\bibliography{reference}